\documentclass[prl,twocolumn,superscriptaddress,citeautoscript]{revtex4-1}

\usepackage[colorlinks]{hyperref}
\usepackage{natbib}
\usepackage{amsmath}    
\usepackage{graphicx}   
\usepackage{verbatim}   
\usepackage{color}      
\usepackage{subfig}  
\usepackage{hyperref}   
\definecolor{orange}{rgb}{1,0.5,0}

\definecolor{darkgreen}{rgb}{0, 0.4, 0.1}

\begin{document}

\title{Sub-femtonewton Force Spectroscopy at the Thermal Limit \\in Liquids}
\author{Lulu Liu}
\email[Email: ]{lululiu@fas.harvard.edu}
\author{Simon Kheifets}
\affiliation{School of Engineering and Applied Sciences, Harvard University, 29 Oxford Street, Cambridge, MA 02138}
\author{Vincent Ginis}
\affiliation{School of Engineering and Applied Sciences, Harvard University, 29 Oxford Street, Cambridge, MA 02138}
\affiliation{Applied Physics, Vrije Universiteit Brussel, Pleinlaan 2, 1050 Brussel, Belgium}
\author{Federico Capasso}
\email[Email: ]{capasso@seas.harvard.edu}
\affiliation{School of Engineering and Applied Sciences, Harvard University, 29 Oxford Street, Cambridge, MA 02138}

\keywords{optical forces, femtonewton, force spectroscopy, thermal limit}

\begin{abstract}
We demonstrate thermally-limited force spectroscopy using a probe formed by a dielectric microsphere optically trapped in water near a dielectric surface. We achieve force resolution below 1 fN in 100 s, corresponding to a 2 \AA ~RMS displacement of the probe. Our measurement combines a calibrated evanescent wave particle tracking technique and a lock-in detection method. We demonstrate the accuracy of our method by measurement of the height-dependent force exerted on the probe by an evanescent wave, the results of which are in agreement with Mie theory calculations.
\end{abstract}

\maketitle

\paragraph{Introduction.}
Force spectroscopy on the nanoscale is important in studies of fundamental physics \cite{Tkachenko2014,Bueno2004,Munday2009,Bliokh2013}, chemistry \cite{helden08,b99,Helden2015,Israelachvili1996,Helden2003}, and molecular biology \cite{Neuman2008,Florin1994,Kellermayer1997,Abbondanzieri2005,sa04,Alonso2003} and in the design of micro- and nano- opto-mechanical systems \cite{marago2013,Woolf2009,Wiederhecker2009}. In such systems, forces are typically measured via the position read-out of a mechanically compliant probe such as a mechanical cantilever or an optically trapped particle \cite{Mueller2015,Cappella1999,Grier2003}. The smallest measurable force is fundamentally limited by the presence of a fluctuating background force exerted on the probe by its thermal environment. For a probe coupled to a fluid at temperature $T$, with a drag coefficient $\gamma$, the fluctuation-dissipation theorem predicts the (single sided) power spectral density (PSD) for fluctuations of the thermal force to be~\cite{Gittes1998b,Kubo1966},
\begin{align}
\label{eq:psd} 
S^\text{th}_F(\omega) = 4 \gamma k_B T,
\end{align} 
where $k_B$ is the Boltzmann constant, and $\gamma$ relates the dissipative drag force, $F_d$, to the velocity $v$ of the probe according to $F_d = -\gamma v$. For a continuous force measurement with bandwidth $\Delta \omega$ at a frequency $\omega_0$, thermal fluctuations add a stochastic contribution, with a standard deviation of 
\begin{align}
\sigma_{th}=\sqrt{S^\text{th}_F(\omega_0) \Delta \omega}.
\label{eq:limit}
\end{align}

High sensitivity measurements require conditions which minimize $S^\text{th}_F$. Damping can be reduced by operating in high vacuum and using high-Q (low internal damping) oscillators, while $T$ can be minimized by cooling the system to cryogenic levels~\cite{Geraci2010,Giessibl2003}. Such measures, however, cannot be implemented for aqueous systems, in which the temperature is restricted to be near room temperature, and the viscosity of water results in damping orders of magnitude greater than in air. Force detection in an aqueous environment, however, is vital for measurement of biological systems, investigation of forces mediated by fluids~\cite{Munday2009} or to eliminate capillary effects present during measurements in atmosphere \cite{Cappella1999}. 

Optically trapped microspheres have an advantage over AFM cantilevers for measurements in water due to their small size. A 1 $\mu$m diameter sphere in water has a damping coefficient below $10^{-8}$ kg/s, corresponding to a thermally-limited sensitivity of $\sim$10 fN/$\sqrt{\mathrm{Hz}}$, much smaller than that of a typical AFM tip, especially when it is near a surface~\cite{Gittes1998b}. However, attaining thermally limited sensitivity in DC force measurements using trapped microspheres is technically difficult due to the presence of $1/f$ noise sources. Significant effort is necessary to eliminate such noise sources, including the trapping of multiple microspheres and enclosing the experiment in a box filled with helium~\cite{Abbondanzieri2005}.

Forces as small as 20~fN (with averaging times of at least 10 minutes) have been measured using a method known as photonic force microscopy (PFM)~\cite{Helden2015,Florin1997,Rohrbach2005}, in which the force is determined from the statistics of the thermal motion of the particle. However, PFM involves a trade-off between position resolution and force sensitivity: the thermally-limited force sensitivity of PFM scales with $n^{3/2}$, and approaches Eq.~\ref{eq:limit} for $n=1$~\cite{supplementary}, where $n$ is the number of points along which the thermally sampled potential is differentiated.

In this work, we measure height-dependent forces exerted by an evanescent field on an optically trapped probe near a dielectric surface, in water. We achieve thermally limited force measurement, and avoid $1/f$ noise by using a lock-in detection method in which the external force is modulated during the measurement. The position of the particle is read out using a previously described calibrated total internal reflection microscopy (TIRM)~\cite{pr90} method, enabled by the addition of an anti-reflection (AR) coating to the surface which eliminates standing-wave modulation of the optical trap~\cite{Liu2014}. Probe height is scanned by displacing the trapping objective (the trap focus) relative to the chamber. Each force measurement corresponds to a spatial average over the trapping volume, giving a height resolution of about 10~nm.

%
\paragraph{Force measurement with a compliant probe.}

The linear response of a compliant probe to an external force can be described by its mechanical susceptibility $\chi(\omega)$, which, for an optically trapped sphere in liquid, can be approximated as:~\cite{Gittes1998b}
\begin{align}
\chi(\omega) = \frac{1}{\kappa-m\omega^2+i\gamma\omega},
\label{eq:susc}
\end{align}

where $m$ is the mass of the probe, $\kappa$ is the spring constant of the harmonic potential and $\gamma$ is the drag coefficient. For a sphere of radius $r$, in a fluid with viscosity $\eta$, the drag is given by Stokes law, $\gamma_\infty = 6\pi \eta r$. The presence of a planar boundary near the sphere results in an increased, height-dependent drag coefficient and requires a modification to Stokes law~\cite{supplementary}. Knowledge of $\chi(\omega)$ allows one to infer the force on such a probe from a measurement of its displacement, according to the relation $\tilde{F}(\omega)=\tilde{x}(\omega)/\chi(\omega)$. For time scales longer than the momentum relaxation time, $\tau_p = m/\gamma$ \cite{kheifets2014}, the mass term in Eq.~\ref{eq:susc} is negligible, and the forces are dominated by viscous damping and the restoring force of the harmonic trap. This is the case in our experiment, where $\tau_p$ is below 1 $\mu$s.

The magnitude of fluctuations of the thermal force, given by Eq. (\ref{eq:psd}), is determined only by the dissipative component of the mechanical response, and is spectrally flat. A series of repeated time-averaged measurements of duration $t_{\text{m}}$, of a constant force, with such background noise, will have a variance of $S_F^\text{tot}(0)/t_{\text{m}}$~\cite{supplementary}. In practice, however, measurement noise at low frequencies is typically dominated by additional $1/f$ noise sources, the presence of which causes the variance of repeated measurements to diverge.

The effect of low frequency noise can be reduced by modulation of the driving force at a constant frequency $\omega_0$, at which the noise is dominated by thermal noise. In order to extract the signal from the noise, a lock-in algorithm is used. The lock-in signal is generated by time-averaging of the product of the position measurement $x(t)$ and a normalized reference signal $r(t)$, proportional to the external force. Repeated measurements using this method will have a variance of $S_x(\omega_0)/t_{\text{m}}$~\cite{supplementary}, allowing the measurement to be shifted to a spectral range where $1/f$ noise is negligible.

\paragraph{Optical force on a sphere in an evanescent field.}

The optical force exerted by an evanescent field on matter has attracted significant attention in recent decades\cite{Kawata1992,Sasaki2000,Nieto-2004near}, and though general qualitative agreement has been found between Mie theory and observations, due to instrumental limitations, precise quantitative tests have not been made as of yet.

The force on a dielectric sphere in such an optial environment consists of two terms: a gradient force and a scattering force. The gradient force pulls the particle towards the surface and is proportional to the gradient of the field, whereas both the direction and amplitude of the scattering force depend on the scattering properties of the sphere. In the Rayleigh approximation~\cite{okamoto1999radiation}, i.e., when the size of the particle is much smaller than the wavelength of light ($k_0 a \ll 1$), this scattering force is directed parallel to the surface. In the Mie regime ($k_0 a \ge 1$), on the other hand, the incident field can be scattered in every direction, so the scattering force contains components both parallel and perpendicular to the surface. In our experiments, $k_0 a \approx 1$. Therefore, we analytically evaluate the scattered fields of the dielectric sphere using Mie Theory and calculate the resulting optical force by means of an algebraic combination of the Mie scattering coefficients, following the formalism derived in Ref. \cite{Almaas-1995radiation}.  The details of these calculations are discussed in the Supplemental Materials \cite{supplementary}.

\paragraph{Setup.}

\begin{figure}[t!]%
\centering
\includegraphics[width=3.4in]{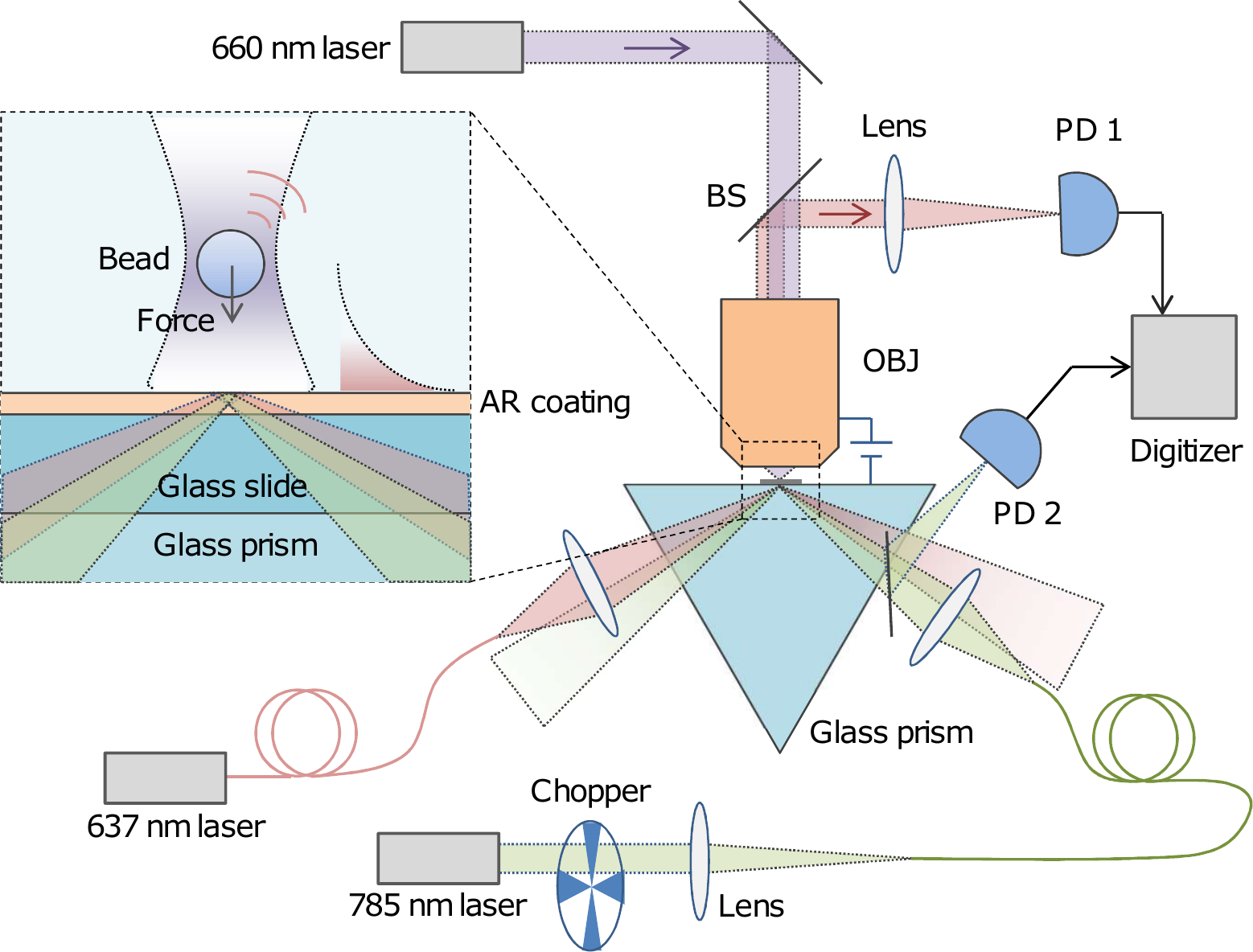}
\captionsetup{justification=raggedright,singlelinecheck=false}
\caption{\footnotesize Setup.  An optical trap is formed by a 660~nm laser beam focused by a high-NA water immersion objective (OBJ). The trap confines polystyrene microspheres in water near an AR coated glass surface. The bead-surface separation is controlled by vertical translation of the objective. A low-intensity evanescent wave used for detection is produced by total internal reflection of a 637 nm (probe) beam. Probe light scattered by the particle is collected by the objective and sent to a low-noise photodiode (PD1). A 660 nm notch filter and two 637 nm bandpass filters placed in front of PD1 eliminate spurious signals from scattered trap and probe beams.  A second, high intensity, 785 nm (pump) evanescent wave exerts a downward gradient optical force on the particle. A chopper modulates the intensity of the pump beam, which is monitored by PD2.  This signal and the detection signal are digitized and recorded by a computer for lock-in signal processing.}%
\label{fig:setup}%
\end{figure}

A simplified schematic of the experimental setup is shown in Fig.~\ref{fig:setup}. The experiment is performed in a 25~$\mu$m deep microfluidic chamber with one inlet and one outlet with AR coating on the lower surface. Microsphere suspensions are introduced into the chamber by a syringe pump.  




The chamber is mounted on a 60$^\circ$ BK-7 glass prism with refractive-index matching fluid in the interface, providing optical access to the lower water-glass interface of the chamber for the detection and actuation beams at angles of incidence beyond the critical angle.

Trapping is achieved using a single-beam optical tweezer formed by a 10 mW, 660~nm laser (Newport LQC660-110C) focused inside the chamber, from above, by a water-immersion objective (Leica PL APO 63$\times$/1.20 W CORR) with numerical aperture (NA) of 1.2. 

The evanescent detection field is formed by a low power ($< 1$~mW) $p$-polarized 637~nm wavelength beam from a fiber-pigtailed laser (Thorlabs LP637-SF70). This ``probe'' beam reflects at the glass-water interface below the trapping area at an at an angle above critical ($\theta_c = 61.04^\circ$). The spot-size of this detection beam was measured to be about $0.13$ mm$^2$, corresponding to an intensity of 0.75 W/cm$^2$ at the surface~\cite{supplementary}. Probe light scattered by the particle is collected by the objective and coupled into a 60~$\mu$m diameter multi-mode fiber, which defines an aperture with an effective diameter of 4~$\mu$m in the image plane. The fiber output is sent to a high-sensitivity photodetector (New Focus 2151) with a transimpedance gain of 2$\times 10^{10}$ V/A. This signal is used to track the particle's separation from the surface.





Actuation of the microsphere is performed using a second, higher power, $\sim$100 mW, 785~nm ``pump'' laser (Thorlabs LD785-SH300) whose beam enters the other side of the prism. It is focused onto an area of about 0.014~mm$^2$, resulting in typical intensities of several kW/cm$^2$. The pump amplitude is modulated by a chopper upstream of the fiber coupler. A 2\% beam splitter is used to deflect a part of the beam into a monitor photodiode, providing the reference signal for lock-in measurement. 


A capacitive displacement sensor mounted between the objective mount and sample holder is used to implement an active, low frequency feedback loop to stabilize the lens-sample separation, which is controlled by a piezo-actuated translation stage. In the presence of feedback, fluctuations in focal position of the optical trap were reduced from that of order 100~nm in 10~min to below 5~nm in 60~min.


\paragraph{Data acquisition and analysis.}
All data reported in this paper were acquired with optically trapped non-functionalized polystyrene (PS) microbeads suspended in de-ionized water.  Typical trap lifetimes were observed to be several hours, limited by eventual diffusion of additional particles into the trapping volume. Stable trapping enabled multiple repeated acquisitions for the same microsphere. A typical data acquisition procedure for a given trapped bead consists of an initial calibration scan followed by lock-in force measurements for different values of trap position and pump beam intensity. Upon completion of the required force measurements, a 0.1M NaCl aqueous solution is introduced into the fluid chamber for a final calibration scan to reduce surface repulsion due to the electrostatic double-layer in water~\cite{f93,clapp2001direct}.

\begin{figure}[htbp]%
\centering
\includegraphics[width=3.4in]{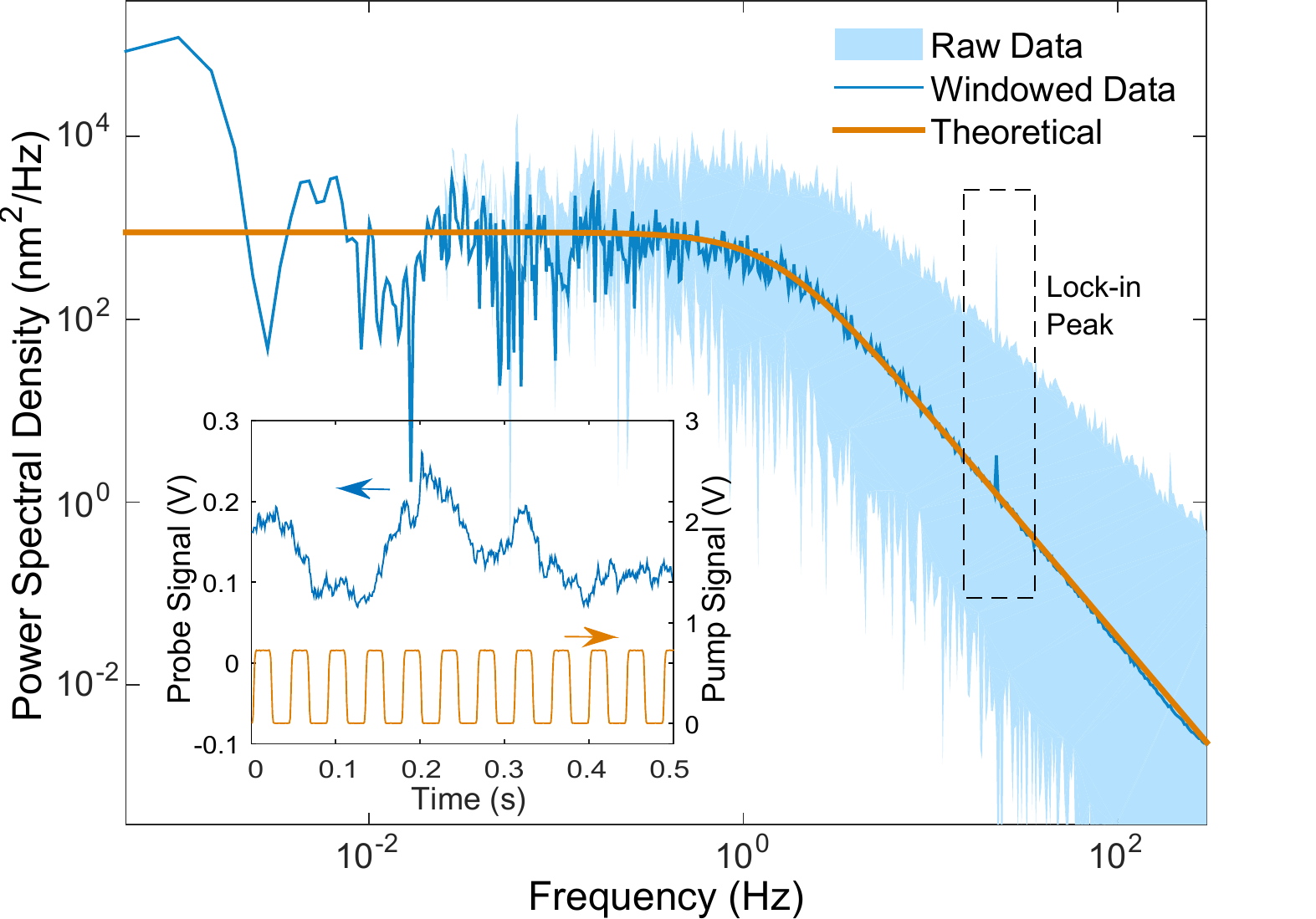}
\captionsetup{justification=raggedright,singlelinecheck=false}
\caption{ \footnotesize Power spectral density (PSD) of the position fluctuations of an optically trapped and optically driven Brownian probe. At low frequencies $1/f$ noise dominates and the PSD deviates from theory. A lock-in peak centered at approximately 23 Hz is visible in the frequency domain though the driven motion is obscured by thermal fluctuations in the time domain, as illustrated by the inset. The lock-in amplitude is determined algorithmically using the pump reference signal.}%
\label{fig:psd}%
\end{figure}


The calibration procedure~\cite{Liu2014} determines the parameters used to convert from the probe signal photodiode voltage $V(z)$ to the absolute distance $z$ between the microsphere and coverslip surface. The trapping objective is lowered in 10~nm steps towards the bottom surface of the coverslip while the average photodiode voltage is recorded for each trap position. The result is fit to the relationship: 
\begin{align}
V(z) = 
\begin{cases}
V_0 e^{-\beta z}+C, &  z \geq 0 \\
V_0, & z < 0
\end{cases}
\end{align}


where $V_0$ is the photodiode voltage (proportional to intensity of scattered light) when the microsphere is in contact with the surface, and $\beta^{-1}$ is the decay length of the intensity of the evanescent field produced by the detection beam~\cite{pr93}.

During force measurements, the particle tracking photodiode signal is recorded simultaneously with that of the pump beam monitor photodiode, which forms the reference signal for the lock-in measurement described above. The output of the lock-in algorithm is the phase and amplitude of the driven oscillation of the probe's position. In order to convert to force, an additional calibration step must be performed to determine the $\chi(\omega)$ in Eq.~(\ref{eq:susc}). The height-dependent susceptibility is found ~\cite{supplementary} using $S_z(\omega,z)$, the PSD of the Brownian motion of the probe acquired at different heights, as exemplified by Figure~\ref{fig:psd}. Additionally, the resulting $\gamma(z)$ is used to determine the diameter of each microsphere by fitting to a near-wall hindered diffusion model~\cite{b61}, with diameter and point of contact as fitting parameters. 


\paragraph{Results.}
%
\begin{figure}[t!]%
\includegraphics[width=3.4in]{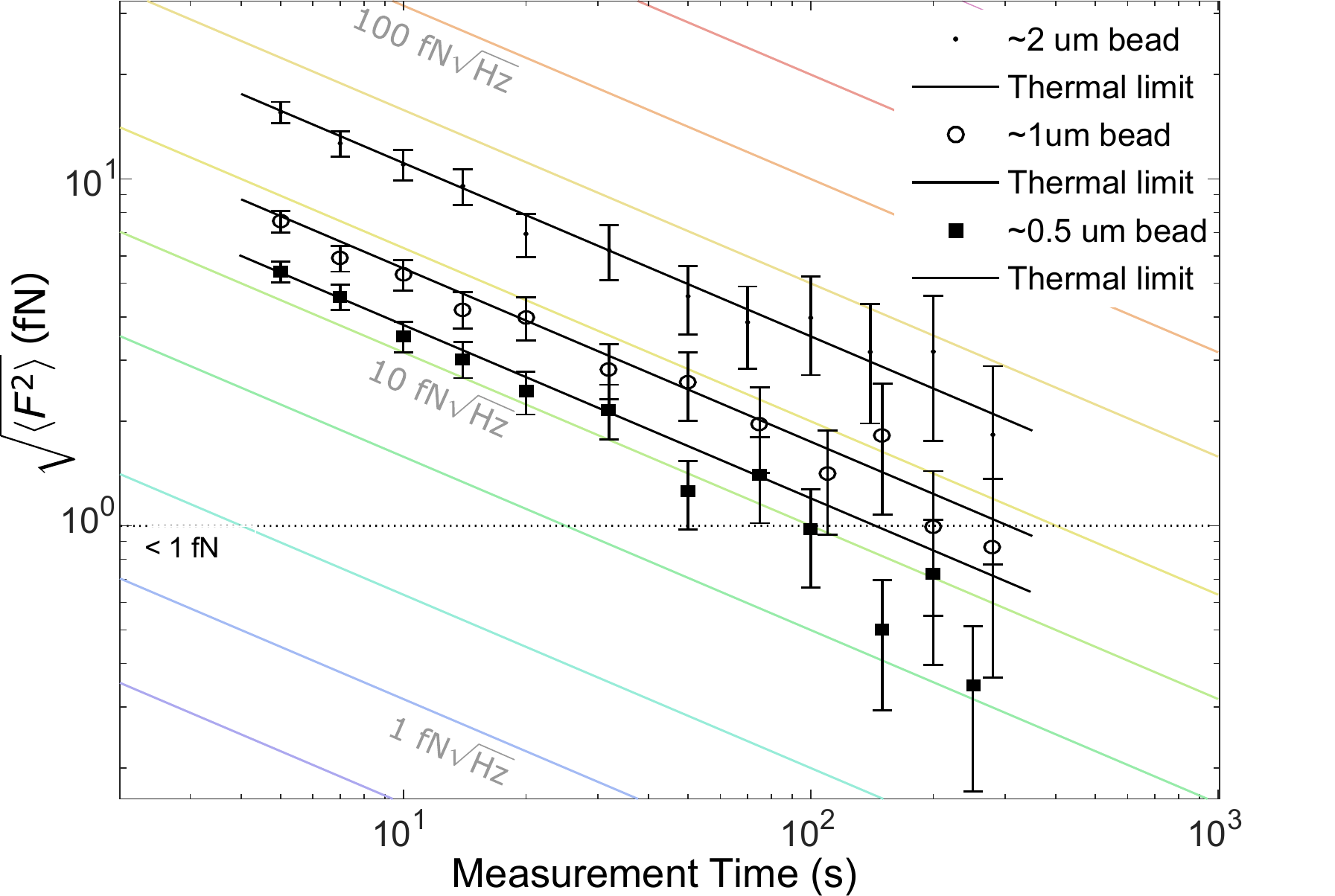}%
\captionsetup{justification=raggedright,singlelinecheck=false}
\caption{ \footnotesize Measured RMS uncertainty in force as a function of measurement time for three beads of different radii.  Each data point is obtained by calculating the standard deviation of a series of repeated independent measurements resampled from a 1000 second time series.  The maximum number of independent measurements used is 25 (for the shortest time points) and minimum is 5 (for the longest time points).  Comparisons are made with the predicted thermal noise limit calculated from measured damping coefficients (solid lines). Contour lines of constant damping strength ($\gamma$) are shown in the background in color.  Dotted line indicates 1~fN level.}%
\label{fig:noise_vs_measurementtime}%
\end{figure}


Figure~\ref{fig:noise_vs_measurementtime} shows the RMS force fluctuations as a function of measurement time (${t_\text{m}}$) for three microspheres with nominal diameters of 500~nm, 1~$\mu$m, and 2~$\mu$m, trapped 700~nm from the surface in the presence of a constant (oscillating), attractive optical force. The solid curves give the prediction of Eq.~(\ref{eq:limit}), and demonstrate that the fluctuations observed in our experimental data are primarily thermally limited. The thermal limit for a smaller particle is itself smaller due to the reduced value for $\gamma$, which is, to first order, proportional to the particle diameter. For measurement times near and above 100 s, the RMS error for the smallest particle drops below 1 fN.




\begin{figure}[htb]%
\centering
\includegraphics[width=3.4in]{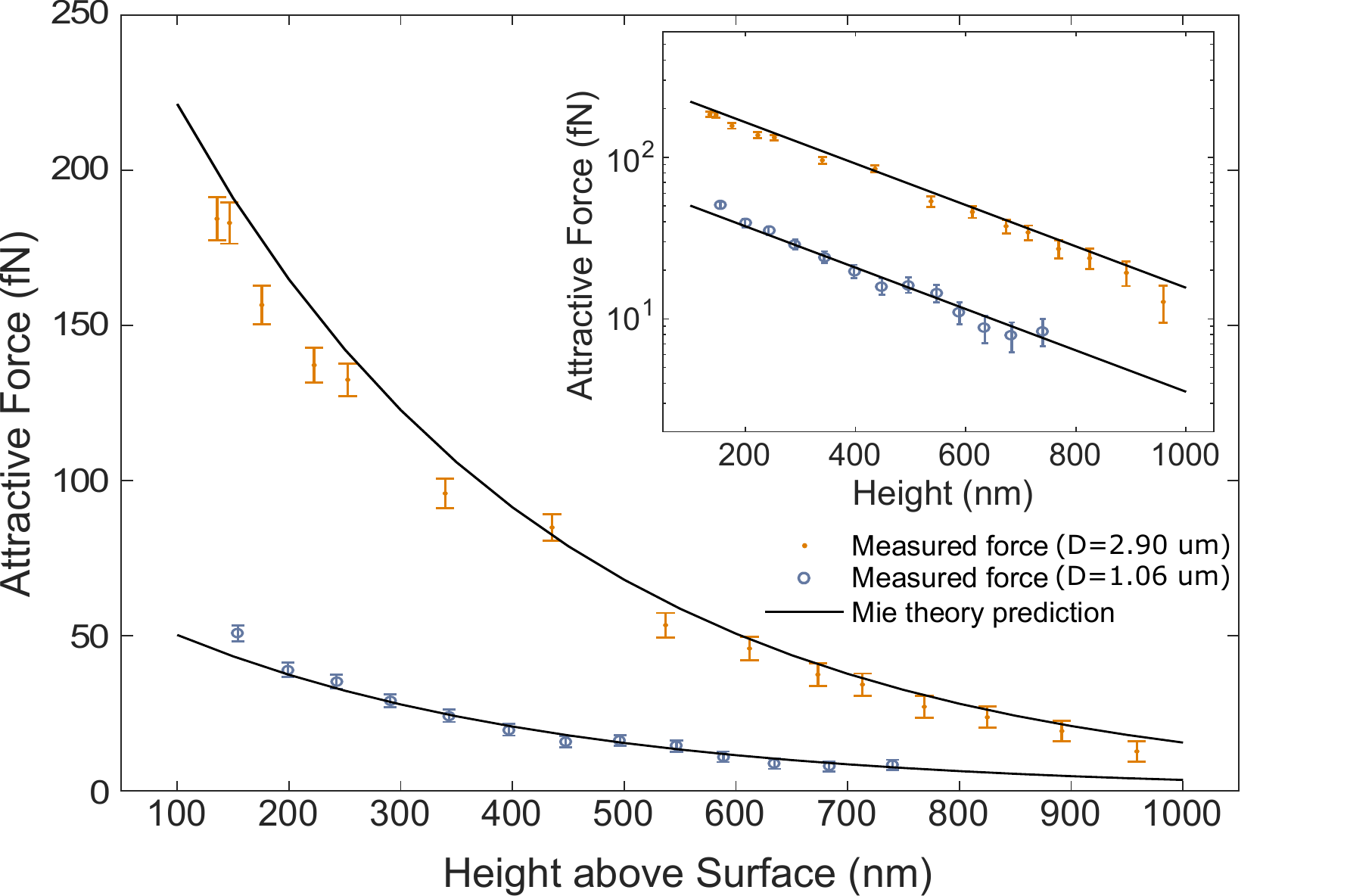}
\captionsetup{justification=raggedright,singlelinecheck=false}
\caption{ \footnotesize Measured optical force on a 2.90 $\mu$m and a 1.06 $\mu$m diameter polystyrene bead in the presence of an evanescent field. The angle of incidence of the pump beam used to generate the optical force was measured to be 62.1$^\circ$ $\pm0.1^\circ$ from normal for both measurements (above the critical angle of 61.04$^\circ$).  Solid lines represent analytical Mie theory prediction of the attractive optical force in the corresponding system.  The inset displays the same data on semilog axes.} %
\label{fig:force_vs_height}%
\end{figure}

Figure \ref{fig:force_vs_height} shows results of measured (attractive) optical force exerted by the near-field of the totally internally reflecting pump beam. Each data point was obtained from a 100 s long measurement at fixed bead-surface separation. The error bars show the thermal limit, confirmed to be equivalent to our force resolution.  The measured force is shown alongside the results of Mie theory calculations (solid curves) based solely on measured parameters.  These parameters are, specifically, the microsphere diameter, the pump laser angle of incidence and field intensity at the surface.  


The microsphere diameters, determined by fitting to the hindered diffusion model, were 1.06 $\pm$ 0.02 $\mu$m and 2.90 $\pm$ 0.06 $\mu$m.  The pump angle of incidence, $\theta_p$, is determined by measurement of the decay length of the evanescent field via confocal detection of the scattered light (blocked by a bandpass filter during normal acquisition). It was determined to be 62.1$^\circ$ $\pm$ 0.1$^\circ$. The field intensity at the surface was estimated from measurement of the laser power incident on the prism, the beam profile at the lower surface of the chamber, and the reflection and propagation losses through the prism. The resulting peak intensities are (1.49 $\pm$ 0.16) and (1.66 $\pm$ 0.17) kW/cm$^2$ for the larger and smaller beads, respectively. Uncertainty in the power meter measurement and beam waist determination are the main sources of error. Detailed calculations are in Supplemental Materials \cite{supplementary}.


Data for the 1.06 $\mu$m microsphere are a near exact match with theory and results for the 2.90 $\mu$m microsphere are close to predicted as well, scaled by a factor of 0.85, which may be attributed to a small pump beam misalignment. Within uncertainties, our experiment confirms Mie theory predictions, and is, to our knowledge, the first such experiment to find precise quantitative agreement between predicted and measured optical forces from an evanescent field.

%




\paragraph{Discussion.}

We have developed a novel technique for conducting thermal-limited force spectroscopy using optically trapped microspheres. We have used it to measure the height dependence of the attractive force of an evanescent wave on a dielectric microsphere, demonstrating quantitative agreement with Mie theory. We have demonstrated resolution of angstrom-scale probe displacements and the detection of sub-femtonewton optical forces in aqueous environments at room temperature, surpassing the sensitivity (in N/$\sqrt{\text{Hz}}$) of existing techniques by two orders of magnitude~\cite{meiners2000femtonewton}.

The improvement makes accessible many optical phenomena of interest, including measurement of lateral spin momentum forces, chiral-sorting forces~\cite{Bliokh2013,Hayat2015, Wang2014} and attractive surface plasmon forces~\cite{vo06}. Though our demonstrations were limited to optical forces at visible wavelengths, the technique is broadly applicable to near-IR wavelengths suitable for biological samples and to any force which can be externally modulated, that is, turned on and off. This modulation can be achieved, for example, by optical, mechanical, or electrical activation of biological or chemical systems.  That the method is also minimally invasive, contact-free, and can be performed in fluid at room temperature is of particular interest to the biological community.  It may enable exploration of the potentially important role played by femtonewton forces in DNA structure formation and cellular function and motility~\cite{Chen2010,Herling2015,Evans1995}.




\begin{acknowledgments}
We acknowledge the support of NSF GFRP grant number DGE1144152 and the Research Foundation Flanders grant number 12O9115N. All fabrication was done in the Harvard Center for Nanoscale Sciences (CNS) clean room facility.  We kindly thank the groups of Evelyn Hu and David Weitz at Harvard University for shared equipment and lab access, and Alexander Woolf for his contributions.
\end{acknowledgments}

\bibliographystyle{apsrev}
\bibliography{optical_forces_bib,library}

\begin{thebibliography}{47}
\expandafter\ifx\csname natexlab\endcsname\relax\def\natexlab#1{#1}\fi
\expandafter\ifx\csname bibnamefont\endcsname\relax
  \def\bibnamefont#1{#1}\fi
\expandafter\ifx\csname bibfnamefont\endcsname\relax
  \def\bibfnamefont#1{#1}\fi
\expandafter\ifx\csname citenamefont\endcsname\relax
  \def\citenamefont#1{#1}\fi
\expandafter\ifx\csname url\endcsname\relax
  \def\url#1{\texttt{#1}}\fi
\expandafter\ifx\csname urlprefix\endcsname\relax\def\urlprefix{URL }\fi
\providecommand{\bibinfo}[2]{#2}
\providecommand{\eprint}[2][]{\url{#2}}

\bibitem[{\citenamefont{Tkachenko and Brasselet}(2014)}]{Tkachenko2014}
\bibinfo{author}{\bibfnamefont{G.}~\bibnamefont{Tkachenko}} \bibnamefont{and}
  \bibinfo{author}{\bibfnamefont{E.}~\bibnamefont{Brasselet}},
  \bibinfo{journal}{Nature Communications} \textbf{\bibinfo{volume}{5}},
  \bibinfo{pages}{3577} (\bibinfo{year}{2014}), ISSN \bibinfo{issn}{2041-1723},
  \urlprefix\url{http://www.ncbi.nlm.nih.gov/pubmed/24717633}.

\bibitem[{\citenamefont{Rugar et~al.}(2004)\citenamefont{Rugar, Budakian,
  Mamin, and Chui}}]{Bueno2004}
\bibinfo{author}{\bibfnamefont{D.}~\bibnamefont{Rugar}},
  \bibinfo{author}{\bibfnamefont{R.}~\bibnamefont{Budakian}},
  \bibinfo{author}{\bibfnamefont{H.}~\bibnamefont{Mamin}}, \bibnamefont{and}
  \bibinfo{author}{\bibfnamefont{B.}~\bibnamefont{Chui}},
  \bibinfo{journal}{Nature} \textbf{\bibinfo{volume}{430}},
  \bibinfo{pages}{329} (\bibinfo{year}{2004}).

\bibitem[{\citenamefont{Munday et~al.}(2009)\citenamefont{Munday, Capasso, and
  Parsegian}}]{Munday2009}
\bibinfo{author}{\bibfnamefont{J.~N.} \bibnamefont{Munday}},
  \bibinfo{author}{\bibfnamefont{F.}~\bibnamefont{Capasso}}, \bibnamefont{and}
  \bibinfo{author}{\bibfnamefont{V.~A.} \bibnamefont{Parsegian}},
  \bibinfo{journal}{Nature} \textbf{\bibinfo{volume}{457}},
  \bibinfo{pages}{170} (\bibinfo{year}{2009}), ISSN \bibinfo{issn}{0028-0836},
  \urlprefix\url{http://www.nature.com/doifinder/10.1038/nature07610}.

\bibitem[{\citenamefont{Bliokh et~al.}(2013)\citenamefont{Bliokh, Bekshaev, and
  Nori}}]{Bliokh2013}
\bibinfo{author}{\bibfnamefont{K.~Y.} \bibnamefont{Bliokh}},
  \bibinfo{author}{\bibfnamefont{A.~Y.} \bibnamefont{Bekshaev}},
  \bibnamefont{and} \bibinfo{author}{\bibfnamefont{F.}~\bibnamefont{Nori}},
  \bibinfo{journal}{Nature Communications} \textbf{\bibinfo{volume}{5}},
  \bibinfo{pages}{14} (\bibinfo{year}{2013}), ISSN \bibinfo{issn}{2041-1723},
  \eprint{1308.0547},
  \urlprefix\url{http://www.ncbi.nlm.nih.gov/pubmed/24598730$\backslash$nhttp://arxiv.org/abs/1308.0547}.

\bibitem[{\citenamefont{Hertlein et~al.}(2008)\citenamefont{Hertlein, Helden,
  Gambassi, Dietrich, and Bechinger}}]{helden08}
\bibinfo{author}{\bibfnamefont{C.}~\bibnamefont{Hertlein}},
  \bibinfo{author}{\bibfnamefont{L.}~\bibnamefont{Helden}},
  \bibinfo{author}{\bibfnamefont{A.}~\bibnamefont{Gambassi}},
  \bibinfo{author}{\bibfnamefont{S.}~\bibnamefont{Dietrich}}, \bibnamefont{and}
  \bibinfo{author}{\bibfnamefont{C.}~\bibnamefont{Bechinger}},
  \bibinfo{journal}{Nature} \textbf{\bibinfo{volume}{451}},
  \bibinfo{pages}{172} (\bibinfo{year}{2008}).

\bibitem[{\citenamefont{Bevan and Prieve}(1999)}]{b99}
\bibinfo{author}{\bibfnamefont{M.~A.} \bibnamefont{Bevan}} \bibnamefont{and}
  \bibinfo{author}{\bibfnamefont{D.~C.} \bibnamefont{Prieve}},
  \bibinfo{journal}{Langmuir} \textbf{\bibinfo{volume}{15}},
  \bibinfo{pages}{7925} (\bibinfo{year}{1999}).

\bibitem[{\citenamefont{Helden et~al.}(2015)\citenamefont{Helden, Eichhorn, and
  Bechinger}}]{Helden2015}
\bibinfo{author}{\bibfnamefont{L.}~\bibnamefont{Helden}},
  \bibinfo{author}{\bibfnamefont{R.}~\bibnamefont{Eichhorn}}, \bibnamefont{and}
  \bibinfo{author}{\bibfnamefont{C.}~\bibnamefont{Bechinger}},
  \bibinfo{journal}{Soft Matter} \textbf{\bibinfo{volume}{11}},
  \bibinfo{pages}{2379} (\bibinfo{year}{2015}), ISSN \bibinfo{issn}{1744-683X},
  \urlprefix\url{http://xlink.rsc.org/?DOI=C4SM02833C}.

\bibitem[{\citenamefont{Israelachvili and
  Wennerstr{\"{o}}m}(1996)}]{Israelachvili1996}
\bibinfo{author}{\bibfnamefont{J.}~\bibnamefont{Israelachvili}}
  \bibnamefont{and}
  \bibinfo{author}{\bibfnamefont{H.}~\bibnamefont{Wennerstr{\"{o}}m}},
  \bibinfo{journal}{Nature} \textbf{\bibinfo{volume}{379}},
  \bibinfo{pages}{219} (\bibinfo{year}{1996}).

\bibitem[{\citenamefont{Helden et~al.}(2003)\citenamefont{Helden, Roth,
  Koenderink, Leiderer, and Bechinger}}]{Helden2003}
\bibinfo{author}{\bibfnamefont{L.}~\bibnamefont{Helden}},
  \bibinfo{author}{\bibfnamefont{R.}~\bibnamefont{Roth}},
  \bibinfo{author}{\bibfnamefont{G.~H.} \bibnamefont{Koenderink}},
  \bibinfo{author}{\bibfnamefont{P.}~\bibnamefont{Leiderer}}, \bibnamefont{and}
  \bibinfo{author}{\bibfnamefont{C.}~\bibnamefont{Bechinger}},
  \bibinfo{journal}{Physical Review Letters} \textbf{\bibinfo{volume}{90}},
  \bibinfo{pages}{048301} (\bibinfo{year}{2003}), ISSN
  \bibinfo{issn}{0031-9007}.

\bibitem[{\citenamefont{Neuman and Nagy}(2008)}]{Neuman2008}
\bibinfo{author}{\bibfnamefont{K.~K.~C.} \bibnamefont{Neuman}}
  \bibnamefont{and} \bibinfo{author}{\bibfnamefont{A.}~\bibnamefont{Nagy}},
  \bibinfo{journal}{Nature Methods} \textbf{\bibinfo{volume}{5}},
  \bibinfo{pages}{491} (\bibinfo{year}{2008}),
  \urlprefix\url{http://www.ncbi.nlm.nih.gov/pmc/articles/PMC3397402/}.

\bibitem[{\citenamefont{Florin et~al.}(1994)\citenamefont{Florin, Moy, and
  Gaub}}]{Florin1994}
\bibinfo{author}{\bibfnamefont{E.}~\bibnamefont{Florin}},
  \bibinfo{author}{\bibfnamefont{V.}~\bibnamefont{Moy}}, \bibnamefont{and}
  \bibinfo{author}{\bibfnamefont{H.}~\bibnamefont{Gaub}},
  \bibinfo{journal}{Science} \textbf{\bibinfo{volume}{264}},
  \bibinfo{pages}{415} (\bibinfo{year}{1994}),
  \eprint{http://www.sciencemag.org/content/264/5157/415.full.pdf},
  \urlprefix\url{http://www.sciencemag.org/content/264/5157/415.abstract}.

\bibitem[{\citenamefont{Kellermayer et~al.}(1997)\citenamefont{Kellermayer,
  Smith, Granzier, and Bustamante}}]{Kellermayer1997}
\bibinfo{author}{\bibfnamefont{M.~S.} \bibnamefont{Kellermayer}},
  \bibinfo{author}{\bibfnamefont{S.~B.} \bibnamefont{Smith}},
  \bibinfo{author}{\bibfnamefont{H.~L.} \bibnamefont{Granzier}},
  \bibnamefont{and}
  \bibinfo{author}{\bibfnamefont{C.}~\bibnamefont{Bustamante}},
  \bibinfo{journal}{Science} \textbf{\bibinfo{volume}{276}},
  \bibinfo{pages}{1112} (\bibinfo{year}{1997}).

\bibitem[{\citenamefont{Abbondanzieri et~al.}(2005)\citenamefont{Abbondanzieri,
  Greenleaf, Shaevitz, Landick, and Block}}]{Abbondanzieri2005}
\bibinfo{author}{\bibfnamefont{E.~A.} \bibnamefont{Abbondanzieri}},
  \bibinfo{author}{\bibfnamefont{W.~J.} \bibnamefont{Greenleaf}},
  \bibinfo{author}{\bibfnamefont{J.~W.} \bibnamefont{Shaevitz}},
  \bibinfo{author}{\bibfnamefont{R.}~\bibnamefont{Landick}}, \bibnamefont{and}
  \bibinfo{author}{\bibfnamefont{S.~M.} \bibnamefont{Block}},
  \bibinfo{journal}{Nature} \textbf{\bibinfo{volume}{438}},
  \bibinfo{pages}{460} (\bibinfo{year}{2005}), ISSN \bibinfo{issn}{1476-4687},
  \urlprefix\url{http://www.pubmedcentral.nih.gov/articlerender.fcgi?artid=1356566{\&}tool=pmcentrez{\&}rendertype=abstract}.

\bibitem[{\citenamefont{Sarkar et~al.}(2004)\citenamefont{Sarkar, Robertson,
  and Fernandez}}]{sa04}
\bibinfo{author}{\bibfnamefont{A.}~\bibnamefont{Sarkar}},
  \bibinfo{author}{\bibfnamefont{R.~B.} \bibnamefont{Robertson}},
  \bibnamefont{and} \bibinfo{author}{\bibfnamefont{J.~M.}
  \bibnamefont{Fernandez}}, \bibinfo{journal}{Proceedings of the National
  Academy of Sciences} \textbf{\bibinfo{volume}{101}}, \bibinfo{pages}{12882}
  (\bibinfo{year}{2004}).

\bibitem[{\citenamefont{Alonso and Goldmann}(2003)}]{Alonso2003}
\bibinfo{author}{\bibfnamefont{J.~L.} \bibnamefont{Alonso}} \bibnamefont{and}
  \bibinfo{author}{\bibfnamefont{W.~H.} \bibnamefont{Goldmann}},
  \bibinfo{journal}{Life Sciences} \textbf{\bibinfo{volume}{72}},
  \bibinfo{pages}{2553} (\bibinfo{year}{2003}), ISSN \bibinfo{issn}{00243205},
  \urlprefix\url{http://linkinghub.elsevier.com/retrieve/pii/S0024320503001656}.

\bibitem[{\citenamefont{Marag{\`o} et~al.}(2013)\citenamefont{Marag{\`o},
  Jones, Gucciardi, Volpe, and Ferrari}}]{marago2013}
\bibinfo{author}{\bibfnamefont{O.~M.} \bibnamefont{Marag{\`o}}},
  \bibinfo{author}{\bibfnamefont{P.~H.} \bibnamefont{Jones}},
  \bibinfo{author}{\bibfnamefont{P.~G.} \bibnamefont{Gucciardi}},
  \bibinfo{author}{\bibfnamefont{G.}~\bibnamefont{Volpe}}, \bibnamefont{and}
  \bibinfo{author}{\bibfnamefont{A.~C.} \bibnamefont{Ferrari}},
  \bibinfo{journal}{Nature Nanotechnology} \textbf{\bibinfo{volume}{8}},
  \bibinfo{pages}{807} (\bibinfo{year}{2013}).

\bibitem[{\citenamefont{Woolf et~al.}(2009)\citenamefont{Woolf, Loncar, and
  Capasso}}]{Woolf2009}
\bibinfo{author}{\bibfnamefont{D.}~\bibnamefont{Woolf}},
  \bibinfo{author}{\bibfnamefont{M.}~\bibnamefont{Loncar}}, \bibnamefont{and}
  \bibinfo{author}{\bibfnamefont{F.}~\bibnamefont{Capasso}},
  \bibinfo{journal}{Optics Express} \textbf{\bibinfo{volume}{17}},
  \bibinfo{pages}{19996} (\bibinfo{year}{2009}), ISSN
  \bibinfo{issn}{1094-4087}.

\bibitem[{\citenamefont{Wiederhecker et~al.}(2009)\citenamefont{Wiederhecker,
  Chen, Gondarenko, and Lipson}}]{Wiederhecker2009}
\bibinfo{author}{\bibfnamefont{G.~S.} \bibnamefont{Wiederhecker}},
  \bibinfo{author}{\bibfnamefont{L.}~\bibnamefont{Chen}},
  \bibinfo{author}{\bibfnamefont{A.}~\bibnamefont{Gondarenko}},
  \bibnamefont{and} \bibinfo{author}{\bibfnamefont{M.}~\bibnamefont{Lipson}},
  \bibinfo{journal}{Nature} \textbf{\bibinfo{volume}{462}},
  \bibinfo{pages}{633} (\bibinfo{year}{2009}), ISSN \bibinfo{issn}{0028-0836},
  \urlprefix\url{http://www.nature.com/doifinder/10.1038/nature08584}.

\bibitem[{\citenamefont{Mueller et~al.}(2008)\citenamefont{Mueller, Heugel, and
  Wang}}]{Mueller2015}
\bibinfo{author}{\bibfnamefont{F.}~\bibnamefont{Mueller}},
  \bibinfo{author}{\bibfnamefont{S.}~\bibnamefont{Heugel}}, \bibnamefont{and}
  \bibinfo{author}{\bibfnamefont{L.}~\bibnamefont{Wang}},
  \bibinfo{journal}{Optics Letters} \textbf{\bibinfo{volume}{33}},
  \bibinfo{pages}{539} (\bibinfo{year}{2008}).

\bibitem[{\citenamefont{Cappella and Dietler}(1999)}]{Cappella1999}
\bibinfo{author}{\bibfnamefont{B.}~\bibnamefont{Cappella}} \bibnamefont{and}
  \bibinfo{author}{\bibfnamefont{G.}~\bibnamefont{Dietler}},
  \bibinfo{journal}{Surface Science Reports} \textbf{\bibinfo{volume}{34}},
  \bibinfo{pages}{1} (\bibinfo{year}{1999}), ISSN \bibinfo{issn}{01675729}.

\bibitem[{\citenamefont{Grier}(2003)}]{Grier2003}
\bibinfo{author}{\bibfnamefont{D.~G.} \bibnamefont{Grier}},
  \bibinfo{journal}{Nature} \textbf{\bibinfo{volume}{424}},
  \bibinfo{pages}{810} (\bibinfo{year}{2003}), ISSN \bibinfo{issn}{1476-4687},
  \urlprefix\url{http://www.ncbi.nlm.nih.gov/pubmed/12917694}.

\bibitem[{\citenamefont{Gittes and Schmidt}(1998)}]{Gittes1998b}
\bibinfo{author}{\bibfnamefont{F.}~\bibnamefont{Gittes}} \bibnamefont{and}
  \bibinfo{author}{\bibfnamefont{C.~F.} \bibnamefont{Schmidt}},
  \bibinfo{journal}{European Biophysics Journal} \textbf{\bibinfo{volume}{27}},
  \bibinfo{pages}{75} (\bibinfo{year}{1998}), ISSN \bibinfo{issn}{01757571}.

\bibitem[{\citenamefont{Kubo}(1966)}]{Kubo1966}
\bibinfo{author}{\bibfnamefont{R.}~\bibnamefont{Kubo}},
  \bibinfo{journal}{Reports on Progress in Physics}
  \textbf{\bibinfo{volume}{29}}, \bibinfo{pages}{255} (\bibinfo{year}{1966}),
  ISSN \bibinfo{issn}{00344885},
  \urlprefix\url{http://stacks.iop.org/0034-4885/29/i=1/a=306?key=crossref.d8453ffc416cd2064d6ccd38e7c06a41}.

\bibitem[{\citenamefont{Geraci et~al.}(2010)\citenamefont{Geraci, Papp, and
  Kitching}}]{Geraci2010}
\bibinfo{author}{\bibfnamefont{A.}~\bibnamefont{Geraci}},
  \bibinfo{author}{\bibfnamefont{S.}~\bibnamefont{Papp}}, \bibnamefont{and}
  \bibinfo{author}{\bibfnamefont{J.}~\bibnamefont{Kitching}},
  \bibinfo{journal}{Physical Review Letters} \textbf{\bibinfo{volume}{105}},
  \bibinfo{pages}{101101} (\bibinfo{year}{2010}), ISSN
  \bibinfo{issn}{0031-9007},
  \urlprefix\url{http://link.aps.org/doi/10.1103/PhysRevLett.105.101101}.

\bibitem[{\citenamefont{Giessibl}(2003)}]{Giessibl2003}
\bibinfo{author}{\bibfnamefont{F.~J.} \bibnamefont{Giessibl}},
  \bibinfo{journal}{Reviews of Modern Physics} \textbf{\bibinfo{volume}{75}},
  \bibinfo{pages}{949} (\bibinfo{year}{2003}), ISSN \bibinfo{issn}{0034-6861},
  \urlprefix\url{http://link.aps.org/doi/10.1103/RevModPhys.75.949}.

\bibitem[{\citenamefont{Florin et~al.}(1997)\citenamefont{Florin, Pralle,
  H{\"o}rber, and Stelzer}}]{Florin1997}
\bibinfo{author}{\bibfnamefont{E.-L.} \bibnamefont{Florin}},
  \bibinfo{author}{\bibfnamefont{A.}~\bibnamefont{Pralle}},
  \bibinfo{author}{\bibfnamefont{J.~H.} \bibnamefont{H{\"o}rber}},
  \bibnamefont{and} \bibinfo{author}{\bibfnamefont{E.~H.}
  \bibnamefont{Stelzer}}, \bibinfo{journal}{Journal of Structural Biology}
  \textbf{\bibinfo{volume}{119}}, \bibinfo{pages}{202} (\bibinfo{year}{1997}).

\bibitem[{\citenamefont{Rohrbach}(2005)}]{Rohrbach2005}
\bibinfo{author}{\bibfnamefont{A.}~\bibnamefont{Rohrbach}},
  \bibinfo{journal}{Optics Express} \textbf{\bibinfo{volume}{13}},
  \bibinfo{pages}{9695} (\bibinfo{year}{2005}), ISSN \bibinfo{issn}{1094-4087}.

\bibitem[{sup()}]{supplementary}
\bibinfo{note}{See Supplemental Material at
  http://link.aps.org/supplemental/xx/PhysRevLett.xx for a detailed discussion
  of the experimental setup, the measurement algorithm, the Mie theory
  algorithm, and the data analysis.}

\bibitem[{\citenamefont{Prieve et~al.}(1990)\citenamefont{Prieve, Bike, and
  Frej}}]{pr90}
\bibinfo{author}{\bibfnamefont{D.~C.} \bibnamefont{Prieve}},
  \bibinfo{author}{\bibfnamefont{S.~G.} \bibnamefont{Bike}}, \bibnamefont{and}
  \bibinfo{author}{\bibfnamefont{N.~A.} \bibnamefont{Frej}},
  \bibinfo{journal}{Faraday Discussions of the Chemical Society}
  \textbf{\bibinfo{volume}{90}}, \bibinfo{pages}{209} (\bibinfo{year}{1990}).

\bibitem[{\citenamefont{Liu et~al.}(2014)\citenamefont{Liu, Woolf, Rodriguez,
  and Capasso}}]{Liu2014}
\bibinfo{author}{\bibfnamefont{L.}~\bibnamefont{Liu}},
  \bibinfo{author}{\bibfnamefont{A.}~\bibnamefont{Woolf}},
  \bibinfo{author}{\bibfnamefont{A.~W.} \bibnamefont{Rodriguez}},
  \bibnamefont{and} \bibinfo{author}{\bibfnamefont{F.}~\bibnamefont{Capasso}},
  \bibinfo{journal}{Proceedings of the National Academy of Sciences}
  \textbf{\bibinfo{volume}{111}}, \bibinfo{pages}{E5609}
  (\bibinfo{year}{2014}), ISSN \bibinfo{issn}{0027-8424},
  \urlprefix\url{http://www.pnas.org/lookup/doi/10.1073/pnas.1422178112}.

\bibitem[{\citenamefont{Kheifets et~al.}(2014)\citenamefont{Kheifets, Simha,
  Melin, Li, and Raizen}}]{kheifets2014}
\bibinfo{author}{\bibfnamefont{S.}~\bibnamefont{Kheifets}},
  \bibinfo{author}{\bibfnamefont{A.}~\bibnamefont{Simha}},
  \bibinfo{author}{\bibfnamefont{K.}~\bibnamefont{Melin}},
  \bibinfo{author}{\bibfnamefont{T.}~\bibnamefont{Li}}, \bibnamefont{and}
  \bibinfo{author}{\bibfnamefont{M.~G.} \bibnamefont{Raizen}},
  \bibinfo{journal}{science} \textbf{\bibinfo{volume}{343}},
  \bibinfo{pages}{1493} (\bibinfo{year}{2014}).

\bibitem[{\citenamefont{Kawata and Sugiura}(1992)}]{Kawata1992}
\bibinfo{author}{\bibfnamefont{S.}~\bibnamefont{Kawata}} \bibnamefont{and}
  \bibinfo{author}{\bibfnamefont{T.}~\bibnamefont{Sugiura}},
  \bibinfo{journal}{Optics Letters} \textbf{\bibinfo{volume}{17}},
  \bibinfo{pages}{772} (\bibinfo{year}{1992}).

\bibitem[{\citenamefont{Sasaki et~al.}(2000)\citenamefont{Sasaki, Hotta, Wada,
  and Masuhara}}]{Sasaki2000}
\bibinfo{author}{\bibfnamefont{K.}~\bibnamefont{Sasaki}},
  \bibinfo{author}{\bibfnamefont{J.}~\bibnamefont{Hotta}},
  \bibinfo{author}{\bibfnamefont{K.}~\bibnamefont{Wada}}, \bibnamefont{and}
  \bibinfo{author}{\bibfnamefont{H.}~\bibnamefont{Masuhara}},
  \bibinfo{journal}{Optics Letters} \textbf{\bibinfo{volume}{25}},
  \bibinfo{pages}{1385} (\bibinfo{year}{2000}), ISSN \bibinfo{issn}{0146-9592}.

\bibitem[{\citenamefont{Nieto-Vesperinas
  et~al.}(2004)\citenamefont{Nieto-Vesperinas, Chaumet, and
  Rahmani}}]{Nieto-2004near}
\bibinfo{author}{\bibfnamefont{M.}~\bibnamefont{Nieto-Vesperinas}},
  \bibinfo{author}{\bibfnamefont{P.}~\bibnamefont{Chaumet}}, \bibnamefont{and}
  \bibinfo{author}{\bibfnamefont{A.}~\bibnamefont{Rahmani}},
  \bibinfo{journal}{Philosophical Transactions of the Royal Society A}
  \textbf{\bibinfo{volume}{362}}, \bibinfo{pages}{719} (\bibinfo{year}{2004}).

\bibitem[{\citenamefont{Okamoto and Kawata}(1999)}]{okamoto1999radiation}
\bibinfo{author}{\bibfnamefont{K.}~\bibnamefont{Okamoto}} \bibnamefont{and}
  \bibinfo{author}{\bibfnamefont{S.}~\bibnamefont{Kawata}},
  \bibinfo{journal}{Physical Review Letters} \textbf{\bibinfo{volume}{83}},
  \bibinfo{pages}{4534} (\bibinfo{year}{1999}).

\bibitem[{\citenamefont{Almaas and Brevik}(1995)}]{Almaas-1995radiation}
\bibinfo{author}{\bibfnamefont{E.}~\bibnamefont{Almaas}} \bibnamefont{and}
  \bibinfo{author}{\bibfnamefont{I.}~\bibnamefont{Brevik}},
  \bibinfo{journal}{Journal of the Optical Society of America B}
  \textbf{\bibinfo{volume}{12}}, \bibinfo{pages}{2429} (\bibinfo{year}{1995}).

\bibitem[{\citenamefont{Flicker and Bike}(1993)}]{f93}
\bibinfo{author}{\bibfnamefont{S.~G.} \bibnamefont{Flicker}} \bibnamefont{and}
  \bibinfo{author}{\bibfnamefont{S.~G.} \bibnamefont{Bike}},
  \bibinfo{journal}{Langmuir} \textbf{\bibinfo{volume}{9}},
  \bibinfo{pages}{257} (\bibinfo{year}{1993}).

\bibitem[{\citenamefont{Clapp and Dickinson}(2001)}]{clapp2001direct}
\bibinfo{author}{\bibfnamefont{A.~R.} \bibnamefont{Clapp}} \bibnamefont{and}
  \bibinfo{author}{\bibfnamefont{R.~B.} \bibnamefont{Dickinson}},
  \bibinfo{journal}{Langmuir} \textbf{\bibinfo{volume}{17}},
  \bibinfo{pages}{2182} (\bibinfo{year}{2001}).

\bibitem[{\citenamefont{Prieve and Walz}(1993)}]{pr93}
\bibinfo{author}{\bibfnamefont{D.~C.} \bibnamefont{Prieve}} \bibnamefont{and}
  \bibinfo{author}{\bibfnamefont{J.~Y.} \bibnamefont{Walz}},
  \bibinfo{journal}{Applied Optics} \textbf{\bibinfo{volume}{32}},
  \bibinfo{pages}{1629} (\bibinfo{year}{1993}).

\bibitem[{\citenamefont{Brenner}(1961)}]{b61}
\bibinfo{author}{\bibfnamefont{H.}~\bibnamefont{Brenner}},
  \bibinfo{journal}{Chemical Engineering Science}
  \textbf{\bibinfo{volume}{16}}, \bibinfo{pages}{242} (\bibinfo{year}{1961}).

\bibitem[{\citenamefont{Meiners and Quake}(2000)}]{meiners2000femtonewton}
\bibinfo{author}{\bibfnamefont{J.-C.} \bibnamefont{Meiners}} \bibnamefont{and}
  \bibinfo{author}{\bibfnamefont{S.~R.} \bibnamefont{Quake}},
  \bibinfo{journal}{Physical Review Letters} \textbf{\bibinfo{volume}{84}},
  \bibinfo{pages}{5014} (\bibinfo{year}{2000}).

\bibitem[{\citenamefont{Hayat et~al.}(2015)\citenamefont{Hayat, Mueller, and
  Capasso}}]{Hayat2015}
\bibinfo{author}{\bibfnamefont{A.}~\bibnamefont{Hayat}},
  \bibinfo{author}{\bibfnamefont{J.~P.~B.} \bibnamefont{Mueller}},
  \bibnamefont{and} \bibinfo{author}{\bibfnamefont{F.}~\bibnamefont{Capasso}},
  \bibinfo{journal}{Proceedings of the National Academy of Sciences}
  \textbf{\bibinfo{volume}{2015}}, \bibinfo{pages}{201516704}
  (\bibinfo{year}{2015}), ISSN \bibinfo{issn}{0027-8424},
  \urlprefix\url{http://www.pnas.org/lookup/doi/10.1073/pnas.1516704112}.

\bibitem[{\citenamefont{Wang and Chan}(2014)}]{Wang2014}
\bibinfo{author}{\bibfnamefont{S.~B.} \bibnamefont{Wang}} \bibnamefont{and}
  \bibinfo{author}{\bibfnamefont{C.~T.} \bibnamefont{Chan}},
  \bibinfo{journal}{Nature Communications} \textbf{\bibinfo{volume}{5}},
  \bibinfo{pages}{1} (\bibinfo{year}{2014}), ISSN \bibinfo{issn}{2041-1723},
  \urlprefix\url{http://www.nature.com/doifinder/10.1038/ncomms4307}.

\bibitem[{\citenamefont{Volpe et~al.}(2006)\citenamefont{Volpe, Quidant,
  Badenes, and Petrov}}]{vo06}
\bibinfo{author}{\bibfnamefont{G.}~\bibnamefont{Volpe}},
  \bibinfo{author}{\bibfnamefont{R.}~\bibnamefont{Quidant}},
  \bibinfo{author}{\bibfnamefont{G.}~\bibnamefont{Badenes}}, \bibnamefont{and}
  \bibinfo{author}{\bibfnamefont{D.}~\bibnamefont{Petrov}},
  \bibinfo{journal}{Physical Review Letters} \textbf{\bibinfo{volume}{96}},
  \bibinfo{pages}{238101} (\bibinfo{year}{2006}).

\bibitem[{\citenamefont{Chen et~al.}(2010)\citenamefont{Chen, Milstein, and
  Meiners}}]{Chen2010}
\bibinfo{author}{\bibfnamefont{Y.~F.} \bibnamefont{Chen}},
  \bibinfo{author}{\bibfnamefont{J.~N.} \bibnamefont{Milstein}},
  \bibnamefont{and} \bibinfo{author}{\bibfnamefont{J.~C.}
  \bibnamefont{Meiners}}, \bibinfo{journal}{Physical Review Letters}
  \textbf{\bibinfo{volume}{104}}, \bibinfo{pages}{1} (\bibinfo{year}{2010}),
  ISSN \bibinfo{issn}{00319007}.

\bibitem[{\citenamefont{Herling et~al.}(2015)\citenamefont{Herling, Garcia,
  Michaels, Grentz, Dean, Shimanovich, Gang, M{\"{u}}ller, Kav, Terentjev
  et~al.}}]{Herling2015}
\bibinfo{author}{\bibfnamefont{T.~W.} \bibnamefont{Herling}},
  \bibinfo{author}{\bibfnamefont{G.~A.} \bibnamefont{Garcia}},
  \bibinfo{author}{\bibfnamefont{T.~C.~T.} \bibnamefont{Michaels}},
  \bibinfo{author}{\bibfnamefont{W.}~\bibnamefont{Grentz}},
  \bibinfo{author}{\bibfnamefont{J.}~\bibnamefont{Dean}},
  \bibinfo{author}{\bibfnamefont{U.}~\bibnamefont{Shimanovich}},
  \bibinfo{author}{\bibfnamefont{H.}~\bibnamefont{Gang}},
  \bibinfo{author}{\bibfnamefont{T.}~\bibnamefont{M{\"{u}}ller}},
  \bibinfo{author}{\bibfnamefont{B.}~\bibnamefont{Kav}},
  \bibinfo{author}{\bibfnamefont{E.~M.} \bibnamefont{Terentjev}},
  \bibnamefont{et~al.}, \bibinfo{journal}{Proceedings of the National Academy
  of Sciences} \textbf{\bibinfo{volume}{112}}, \bibinfo{pages}{1417326112}
  (\bibinfo{year}{2015}), ISSN \bibinfo{issn}{1091-6490},
  \urlprefix\url{http://www.pnas.org/content/early/2015/07/16/1417326112.abstract?sid=5c24a61b-5bba-45a7-bca1-30102a5175dd}.

\bibitem[{\citenamefont{Evans et~al.}(1995)\citenamefont{Evans, Ritchie, and
  Merkel}}]{Evans1995}
\bibinfo{author}{\bibfnamefont{E.}~\bibnamefont{Evans}},
  \bibinfo{author}{\bibfnamefont{K.}~\bibnamefont{Ritchie}}, \bibnamefont{and}
  \bibinfo{author}{\bibfnamefont{R.}~\bibnamefont{Merkel}},
  \bibinfo{journal}{Biophysical journal} \textbf{\bibinfo{volume}{68}},
  \bibinfo{pages}{2580} (\bibinfo{year}{1995}), ISSN \bibinfo{issn}{0006-3495},
  \urlprefix\url{http://www.pubmedcentral.nih.gov/articlerender.fcgi?artid=1282168{\&}tool=pmcentrez{\&}rendertype=abstract}.

\end{thebibliography}

\end{document}